\documentclass[12pt,aps,prb,floats,graphicx,euscript,amsmath] {revtex4}
\usepackage[cp1251]{inputenc}
\usepackage[dvips]{graphicx}
\DeclareMathAlphabet\EuScript{U}{eus}{m}{n} \SetMathAlphabet\EuScript{bold}{U}{eus}{b}{n}

\def\lapprox{\,\raise0.4ex\hbox{$<$}\kern-0.8em\lower0.7ex\hbox{$\sim$}\,}
\def\gapprox{\,\raise0.4ex\hbox{$>$}\kern-0.8em\lower0.7ex\hbox{$\sim$}\,}
\begin{document}
\bibliographystyle{prsty}
\title{Spin-Wave Relaxation in a Quantum Hall Ferromagnet}

\author{S. Dickmann$^{1}$ and S.L. Artyukhin$^{1,2}$}
\affiliation{$^{1}$Institute for Solid State Physics of RAS, Chernogolovka 142432, Moscow
District, Russia.\\
$^{2}$University of Groningen, Broerstraat 5, 9712 CP Groningen, Netherlands}

\date{\today}

\begin{abstract}
\vspace{0.mm} We study spin wave relaxation in quantum Hall
ferromagnet regimes. Spin-orbit coupling is considered as a factor
determining spin nonconservation, and external random potential as
a cause of energy dissipation making spin-flip processes
irreversible. We compare this relaxation mechanism with other
relaxation channels existing in a quantum Hall ferromagnet.

 \noindent PACS numbers 73.21.Fg, 73.43.Lp, 78.67.De
\end{abstract}
\maketitle

\bibliographystyle{prsty}

{\bf 1.} Last years are characterized by growing interest in spin
relaxation (SR) in low-dimension systems --- first of all, in the
relaxation in quantum dots studied within the projects aimed at
development of a computer employing spin memory. Yet, the
relaxation of an electron spin in lateral quantum dots
manufactured on the basis of two-dimensional (2D)
heterostructures, should be in many respects similar to the SR of
electrons localized in the 2D layer in minima of a smooth random
potential (SRP). In high magnetic fields this single-electron
relaxation corresponds to the situation occurring at low Landau
level (LL) filling: $\nu\ll 1$ or $|\nu\!-\!2n|\ll 1$ ($n$ is an
integer).\cite{di03}

The SR at different filing factors, $\nu\gapprox 1$, has quite different nature representing in
this case a many-electron process. In particular, in a quantum Hall ferromagnet (QHF), i.e. at
$\nu=1,3,...$ or $\nu=1/3,1/5,...$, the SR reduces to the relaxation of lowest collective
excitations, i.e. spin waves.\cite{by81,lo93} The SR observation would thereby be a good tool to
study fundamental collective properties of a strongly correlated 2D electron gas (2DEG).
However, in spite of much recent interest in the SR in a 2DEG, up to now only a handful of
experiments relevant to the SR in a QHF were performed: these are indirect results based on the
linewidth measurements in the electron spin resonance,\cite{do88} and a direct observation where
the photoluminescence dynamics of spin-up and spin-down states was studied.\cite{zh93}
Meanwhile, availability of the new time-resolved technique of photon counting allows us to
believe that new direct experiments on observation of excitations' relaxation in a 2DEG, in
particular of the spin wave relaxation (SWR), will become available in the near future.\cite{ku}

Theoretically the SWR in a QHF was studied in works \onlinecite{di96,di99}. It is worth noting
here that the SWR represents actually not spin dephasing but the energy relaxation due to the
spin-flip process. Indeed, any spin-flip means at least dissipation of the Zeeman energy
$\epsilon_{\rm Z}=|g|\mu_BB$ ($g\approx -0.44$ in a GaAs structure). The latter is a part of the
spin-wave (spin exciton, SE) energy
$$
  E_{\rm sw}=\epsilon_{\rm Z}+{\cal E}_q,   \eqno (1)
$$
where ${\cal E}_q$ is the SE correlation energy depending on the
2D wave vector $q$.\cite{by81,lo93} At variance with the
relaxation channel of Ref. \onlinecite{di96} where electron-phonon
interaction was considered as the mechanism making the relaxation
irreversible, and contrary to the case of Ref. \onlinecite{di99}
where the irreversibility was provided by an inter-spin-exciton
interaction mechanism, we now study smooth disorder field as the
reason causing the energy transform. The SRP thereby determines an
alternative relaxation channel competing with the ones studied
earlier. Another distinction of the present work from Refs.
\onlinecite{di96,di99} consists in the study of not only the
integer QHF (at $\nu\!=\!1,3,...$) but also of the fractional one
($\nu\!=\!1/3,1/5,...$) as well. At the same time we again
consider the spin-orbit coupling (SO) as the cause mixing
different spin states and therefore providing the spin
nonconservation. Actually, various SWR channels coexist in
parallel. We consider the total rate and find crossover regions of
external parameters (magnetic field, temperature, etc.) where one
relaxation channel ceases to be dominant and changes into another.

The SR channel due to SRP was already considered in the integer
quantum Hall ferromagnetic case.\cite{di03,di04} However, studied
in these works instead of the SWR was a specific SR when initially
the total macroscopic spin ${\vec S}$ of the system as a whole is
turned away from the equilibrium direction parallel to ${\vec B}$.
(Relaxation of this Goldstone mode microscopically reduces to
annihilation processes of the so-called zero SEs, having exactly
zero momenta.) Contrary to this case, the spin perturbation
determined by excitation of the spin waves (non-zero SEs)
represents an initial deviation where $\Delta S\!=\!\Delta S_z$,
so that ${\vec S}$ is kept parallel to ${\vec B}$ and the total
symmetry of system remains unchanged.

Concerning the origin of SRP, one should note that it has in the 2D layer the ``direct''
component and the effective one. The former is the SRP determined by charged donors located
outside the spacer. The latter is essential in some kinds of quantum wells, being determined by
spatial fluctuations (in the plane of the layer) of quantum well width. These fluctuations lead
to fluctuations of the size-quantization energy and may be presented as an SRP term in the
single electron Hamiltonian. Both SRP components have approximately the same amplitude
$\Delta\!\sim\! 10\,$K and correlation length $\Lambda\!\sim\! 30-50\,$nm.

\vspace{3mm}

 {\bf 2.} The total
Hamiltonian has form $H_{\mbox{\scriptsize tot}}\!= \!\sum_j\!
H_1^{(j)}\!+\!H_{\mbox{\scriptsize int}}$, where $j$ enumerates
electrons, $H_{\rm int}$ is the $e$-$e$ interaction, and the
single-electron operator is
$$
  H_1=\hbar^2\hat{{\bf
  q}}^2/2m_e^*-\epsilon_{\rm Z}\hat{\sigma}_z/2+H_{SO}+\varphi({\bf r})\,.
  \eqno (2)
$$
In this equation $\varphi({\bf r})$ is the SRP field; the SO Hamiltonian is specified for the
(001) GaAs plane,
$$
  H_{SO}=\alpha\left(\hat{{\bf q}}\times\hat{\mbox{\boldmath $\sigma$}}
  \right)_{\!z}\!+\!
  \beta\left(\vphantom{\left(\hat{{\bf q}}\times\hat{\mbox{\boldmath
  $\sigma$}}
  \right)}\hat{ q}_y\hat{\sigma}_y\!-\!{\hat q}_x\hat{\sigma}_x\right)\,,
                                                            \eqno (3)
$$
presenting a combination of the Rashba term and the crystalline
anisotropy term$\;$ \cite{by84} ($\hat{{\bf q}}=-i{\bf
\nabla}+e{\bf A}/c\hbar$ is a 2D operator, $\sigma_{x,y,z}$ are
the Pauli matrices). If the SRP is assumed to be Gaussian, then it
is defined by the correlator $K({\bf r})=\langle \varphi({\bf
r})\varphi(0)\rangle$. By choosing $\langle \varphi({\bf
r})\rangle=0$, in terms of the correlation length $\Lambda$ and
the LL width $\Delta$ the correlator is
$$
  K({\bf r})=\Delta^2\exp{(-r^2/\Lambda^2)}\,.   \eqno (4)
$$

We first find the bare single-electron basis diagonalizing the
Hamiltonian (2) without the SRP field. To within the leading order
in the $H_{SO}$ terms we obtain
$$
\begin{array}{l}
\Psi_{pa}=
  \left( {\psi_{n p}\atop v\sqrt{n\!+\!1}\psi_{n\!+\!1\,p}
  +iu\sqrt{n}\psi_{n\!-\!1\,p}}\right),\qquad{}\qquad{}\qquad{}\vspace{2mm}\\
  {}\qquad{}\qquad{}\qquad{}\Psi_{pb}=\left({-v\sqrt{n}
  \psi_{n\!-\!1\,p}
  +iu\sqrt{n\!+\!1}\psi_{n\!+\!1\,p}\atop
  \psi_{n p}}\right)
\end{array}\vspace{-1mm}
 \eqno(5)
$$
Here $\psi_{n p}$ is the electron wave function in the Landau
gauge, $n$ is the number of the half-filled LL in the odd-integer
quantum Hall regime, i.e. in the $\nu\!=\!2n\!+\!1$ case.
Otherwise, if $\nu\!\leq\!1$, we set $n\!=\!0$. $u$ and $v$ are
small dimensionless parameters:
$u=\beta\sqrt{2}/\l_B\hbar\omega_c$ and
$v=\alpha\sqrt{2}/\l_B\hbar\omega_c$ ($\omega_c$ and $l_B$ are the
cyclotron frequency and the magnetic length, respectively). The
single-electron states thus cease to be purely spin states but
acquire a chirality $a$ or $b$. The spin flip corresponds thereby
to the $a\to b$ process now.

By analogy with previous works$\,$\cite{di03,di96,di99,di04} (see also Ref. \onlinecite{dz83})
we define the SE creation operator
$$
  {\cal Q}_{ab\,{\bf q}}^{\dag}=\frac{1}{\sqrt{ N_{\phi}}}\sum_{p}\,
  e^{-iq_x p}
  b_{p+\frac{q_y}{2}}^{\dag}\,a_{p-\frac{q_y}{2}}\,, \eqno (6)
$$
where $a_p$ and $b_p$ are the Fermi annihilation operators corresponding to states (5), $N_\phi$
is the LL degeneracy number. In Eq. (6) and everywhere below we measure wave vector $q$ in the
$1/l_B$ units. If the ratio $r_{\rm c}\!=\!(\alpha e^2/\kappa l_B)/\hbar\omega_c$ is considered
to be small ($\alpha\!<\!1$ is the averaged formfactor which appears due to finiteness of the
layer thickness), and the SRP and SO terms in Eq. (2) are ignored, then the operator (6) acting
on the ground state in the {\it odd-integer quantum Hall regime} yields the {\it eigen state} of
the total Hamiltonian: namely, $[H_{\rm tot},{\cal Q}_{ab\,{\bf
q}}^{\dag}]|0\rangle\!=\!(\epsilon_{\rm Z}\!+\!{\cal E}_q){\cal Q}_{ab\,{\bf
q}}^{\dag}|0\rangle$, where $|{\rm
0}\rangle\!=\!|\overbrace{\uparrow,\uparrow,...\uparrow}^{\displaystyle{\vspace{-15mm}
\mbox{\tiny{$\;N_\phi$}}}}\,\rangle$. This basic property of the exciton state, ${\cal
Q}_{ab\,{\bf q}}^{\dag}|0\rangle$, is the asymptotically exact one to the first order in $r_{\rm
c}$.

Now consider corrections arising due to the $H_{SO}$ terms. When
presented in terms of basis states (5), spin operators $\int
\Psi^{\dag}{\hat {\bf S}}^2\Psi d^2{\bf r}$ and $\int
\Psi^{\dag}{\hat { S}}_z\Psi d^2{\bf r}$ [where
$\Psi\!=\!\sum_p(a_p\Psi_{pa}\!+\!b_p\Psi_{pb})$] preserve
invariant form up to the second order in $u$ and $v$. However, the
interaction Hamiltonian ${ H}_{\rm int}=\frac{1}{2}\int\! d{\bf
r}_1d{\bf r}_2\,{\Psi}^\dag({\bf r}_2){ \Psi}^\dag({\bf
r}_1)U({\bf r}_1\!-\!{\bf r}_2){ \Psi}({\bf r}_1){\Psi}({\bf
r}_2)$ acquires proportional to $u$ and $v$ terms which correspond
to creation and annihilation of SEs in the system. It is exactly
these terms that lead to the ``coalescence'' channel of the
SWR.\cite{di99} In the present work we study another relaxation
channel. Therefore, neglecting this SO corrections to ${\hat
H}_{\rm int}$, we focus on the SRP term. Calculating $\int
\Psi^{\dag}\varphi({\bf r})\Psi d^2{\bf r}$,  we get the terms
responsible for a spin-flip:
$$
  \hat{\varphi}=N_{\phi}^{1/2}l_B\sum_{\bf q}
  \overline{\varphi}
  ({\bf q})\left(iuq_+-vq_-\right)
  {\cal Q}_{\bf q}+\mbox{H.c.}  \vspace{-2mm}
  \eqno (7)
$$
(it is assumed here that $q\!\ll\! 1$). $\overline{\varphi}({\bf
q})$ is the Fourier component [i.e. ${}\!\varphi\!=\!\sum_{\bf
q}\!\overline{\varphi}({\bf q})e^{i{\bf qr}}\!{}$], and
${}\!q_{\pm}\!=\!\mp i(q_x\!\pm\! iq_y)\!/\!\sqrt{2}$.

At variance with integer QHF, the use of the excitonic basis
${\cal Q}_{ab\,{\bf q}}^{\dag}|0\rangle$ presents only a {\it
model approach} in the case of {\it fractional quantum Hall
regime}. Generally, spin-flip excitations within the same Landau
level might be many-particle rather than two-particle excitations
at fractional filling because the same change of the spin numbers
$\delta S\!=\!\delta S_z\!=\!-1$ may be achieved with
participation of arbitrary number of intra-spin-sublevel
excitations (charge-density waves). These waves are generated by
the operator ${\cal A}^\dag_{\bf q}\!=\!N_{\phi}^{-1/2}{\cal
Q}_{aa{\bf q}}^\dag$ acting on the ground state
 $|{\rm
0}\rangle\!=\!|\overbrace{\uparrow,..\uparrow,..\uparrow}^{\displaystyle{\vspace{-15mm}
\mbox{\tiny{$\;\nu N_\phi$}}}}\,\rangle$.\cite{gi86} It is trivial in the case of integer $\nu$
(${\cal A}^\dag_{\bf q}|{\rm 0}\rangle\!=\!\delta_{{\bf q},\,0}|{\rm 0}\rangle$); however,
states of the ${\cal Q}_{ab\,{\bf q}_1}^{\dag}{\cal A}^\dag_{{\bf q}_2}{\cal A}^\dag_{{\bf
q}_3}...|0\rangle$ type might constitute a basis set if one studies a spin-flip at fractional
$\nu$. On the other hand, a comprehensive phenomenological analysis$\,$\cite{gi86,lo93} suggests
that even the spin-flip basis reduced to single-mode (single-exciton) states would be quite
appropriate, at least for lowest-energy excitations in the case of fractional QHF. This
single-mode approach is indirectly substantiated by the fact that the charge-density wave has a
Coulomb gap$\,$\cite{gi86} which is well larger than the Zeeman gap $\epsilon_{\rm Z}$. Hence
for a fractional QHF, just as in Ref. \onlinecite{lo93}, we will consider the only state ${\cal
Q}_{ab\,{\bf q}}^{\dag}|0\rangle$ to describe the spin-flip excitation. The commutation algebra
for operators ${\cal Q}_{ab\,{\bf q}}^{\dag}$, ${\cal A}^\dag_{{\bf q}'}$ and ${\cal
B}^\dag_{{\bf q}''}\!=\!N_{\phi}^{-1/2}{\cal Q}_{bb{\bf q}''}^\dag$ is certainly the same as for
integer filling,\cite{di96,di99,di04}. However, a difference arises in the calculation of
expectation  $\langle 0|{\cal A}_{{\bf q}}{\cal A}^\dag_{{\bf q}'}|0\rangle$ which is needful
for the following. This value is simply $\delta_{{\bf q},0}\delta_{{\bf q}'\!,0}$ at integer
filling, but at $\nu\!<\!1\,$ it is expressed in terms of the two-particle correlation function
$g(r)$ calculated for the ground state:
$$
  \langle 0|{\cal A}_{{\bf q}}{\cal A}^\dag_{{\bf
  q}'}|0\rangle=\displaystyle{\frac{\nu}{N_\phi}}\left[2\pi\nu\overline{g}(q)e^{q^2/2}\!+\!1\right]\delta_{{\bf
  q}'\!,\,{\bf q}}\,. \eqno (8)
$$
Here $\overline{g}(q)\!=\!\frac{1}{(2\pi)^2}\int\! g(r)e^{-i{\bf
qr}}d^2r$ is the Fourier component. Function $g(r)$ is well known,
e.g., in the case of Laughlin's state.\cite{gi86,gi84} If the
ground state is presented in terms of the Hartree-Fock model, we
get the expression $2\pi
\overline{g}\!=\!\left(N_\phi\delta_{q\!,\,0}\!-\!e^{-q^2/2}\right)$
which does not depend on $\nu$. Besides, at odd-integer filling
factors this Hartree-Fock expression becomes Fourier component of
the {\it exact} correlation function. In the latter case one
should also make the substitution $\nu\!\to\nu\!-\!2n$ in Eq. (8),
i.e. formally set $\nu\!=\!1$ there.

\vspace{3mm}

{\bf 3.} The operator (7) obviously does not conserve the number of SEs. However, if the SWR is
governed by this operator, the corresponding problem can not be solved in terms of a
single-exciton study. Indeed, the SE interaction with the SRP incorporates the energy
$U_{\mbox{{\scriptsize x-}{\tiny SRP}}}\!\sim\! ql_B\Delta/\Lambda$ (the SE possesses the dipole
momentum $el_B[{\bf q}\!\times\!{\hat z}]$)$\,$\cite{by81}. The SE momentum is estimated from
the condition ${\cal E}_q\lapprox T$, and we therefore find that $U_{\mbox{{\scriptsize
x-}{\tiny SRP}}}\!\ll\!\epsilon_{\rm Z},\,T$. Due to this inequality, the energy of annihilating
exciton can not be transformed to anywhere. By analogy with Ref. \onlinecite{di99}, we study a
coalescence process where initial double-exciton state $|i\rangle\!=\!{\cal Q}_{ab\,{\bf
q}_{\!1}}^{\dag}\!{\cal Q}_{ab\,{\bf q}_2}^{\dag}|0\rangle$ transforms to final single-exciton
state $|f\rangle\!=\!{\cal Q}_{ab\,{\bf q}'}^{\dag}|0\rangle$ having the combined energy:
$$
\epsilon_{\rm Z}+{\cal E}_{q'}=2\epsilon_{\rm Z}+{\cal E}_{q_{1}}+{\cal E}_{q_2}  \eqno (9)
$$
(c.f. also the Auger magnetoplasma relaxation considered in Ref.
\onlinecite{dile99}). At the same time, contrary to Ref.
\onlinecite{di99}, there is no momentum conservation in this SWR
channel. Thus the phase volume where the $X_{{\bf
q}_1}\!+\!X_{{\bf q}_2}\!\to\!X_{{\bf q}'}$ transition is possible
turns out to be much larger than that in the coalescence process
of Ref. \onlinecite{di99}. This transition is governed by the
Fermi golden rule probability: $w_{fi}=(2\pi/\hbar)|{\cal
M}_{fi}|^2\delta(E_f-E_i)$, and our immediate task is to calculate
the matrix element ${\cal M}_{fi}\!=\!\nu^{-3/2}\langle
f|\hat{\varphi}|i\rangle$. (The factor $\nu^{-3/2}$ appears due to
the normalization since norms of the $|i\rangle$ and $|f\rangle$
states are $\nu^2$ and $\nu$, respectively.)

We perform the calculation for relevant values of momenta
$q_1,q_2,q'\ll\!1$ which satisfy the conditions ${\cal
E}_{q_{\!1}},{\cal E}_{q_2}\lapprox T\lapprox 1\,$K. (These
inequalities correspond to $q_1,q_2,q'\ll\!1/l_B$ in usual
dimensional units). By employing exciton-operators' commutation
rules$\,$\cite{di96} and evident identities ${\cal Q}_{ab\,{\bf
q}}|0\rangle\!\equiv\!{\cal B}_{{\bf q}}|0\rangle\!\equiv\!0$ and
$\langle 0| {\cal A}_{{\bf q}}|0\rangle\!\equiv\!\nu$, we obtain
with the help of Eqs. (7)-(8) that
$$
{}\!{}\!{}\!{}\!{\cal M}_{fi}({\bf q}_1, {\bf q}_2,{\bf
q}')\!=\!{}\frac{2\pi\nu^{1/2}}{N_\phi^{1/2}}\left[\!\sum_{j=1}^2\overline{g}(\!|{\bf
q}_j\!-\!{\bf q}'\!|)e^{({\bf q}_j\!-\!{\bf q}')^2\!/2}\right]\!\sum_{\bf
q}\overline{\varphi}({\bf q})\!\left(iu{q}_+\!-\!v{q}_-\right)\!\delta_{{\bf q}_1\!+\!{\bf
q}_2\!,\,{\bf q}\!+\!{\bf q}'}.
  \eqno (10)
$$
Besides, within our approximation, $\overline{g}(q)e^{q^2/2}$ should be replaced with
$\left.\overline{g}(q)e^{q^2/2}\right|_{q\to 0}$. The latter quantity is equal to $-1/2\pi$ in
the Hartree-Fock approach or $-1/2\pi\nu$ when calculated in the case of Laughlin's ground state
describing the fractional QHF. So, for $\nu\!=\!1,1/3,1/5,...$, replacing the terms in square
brackets with $-1/\pi\nu$, we obtain a simple result:
$$
\left|{\cal M}_{fi}({\bf q}_1, {\bf q}_2,{\bf q}')\right|^2\!=\!{}4\pi\overline{K}({q})
\left.\frac{q^2(u^2\!+\!v^2)}{\nu N_\phi^{2}}\right|_{{\bf q}\!={\bf q}\!{}_1\!+{\bf
q}{}_2\!-{\bf q}'}. \eqno (11)
$$
It is used that the squared modulus of $\overline{\varphi}({\bf q})$ may be expressed in terms
of Fourier component of the correlator (4): $|\overline{\varphi}({\bf q})|^2\!=\!2\pi
K(q)/\!N_\phi$. In the Hartree-Fock model the expression (11) should be multiplied by $\nu^2$;
therefore the calculated relaxation rate would be by a factor of $\nu^2$ slower. Notice also
that if $\nu\!=\!3,5,...$, one should formally set $\nu\!=\!1$ in Eqs. (10) and (11).

The SWR rate is defined as the difference between the fluxes of
annihilating and created SEs. We assume that the thermodynamic
equilibrium in the system of spin waves is established much faster
than the spin-flip processes occur so that the rate is
$$
  R\!=\!\frac{1}{2}\!\!\sum_{{\bf q}_{\!1}\!,\,{\bf q}_2\!,\,{\bf q}'}\!\!\frac{2\pi}{\hbar}\left|{\cal M}_{fi}
  (\mbox{\boldmath $q$}_1,\mbox{\boldmath $q$}_2,{\bf
q}')\right|^2 \delta\!\left(E_1\!+\!E_2\!-\!E'\right)
  \left[n_1n_2\left(1+n'\right)-n'\left(1+n_1\right)
  \left(1+n_2\right)\right].                         \eqno(12)
$$
The notations used here are $E_i=\epsilon_{\rm Z}\!+\!{\cal E}_{q_i},\:\; n_i\!=\!n(E_i)\:\;$
($i=1,\,2)\:$ and ${}\:E'\!=\! \epsilon_{\rm Z}\!+\!{\cal E}_{q'},\:\; n'\!=\!n(E')$, where the
Bose distribution function is $n(E)=1/({e^{(E\!-\!\mu)/T}\!-\!1})$. The rate $R$ is completely
determined by Eqs. (11)-(12) and is a function of parameters $B$, $T$, and of the total number
of SWs in the system: $N_{\rm x}\!=\sum_{\bf q}n(\epsilon_{\rm Z}\!+\!{\cal E}_q)$. In our case,
when temperature is rather low, we can certainly use quadratic approximation for the ``kinetic''
exciton energy: ${\cal E}_q\!\approx\!q^2/2M_{\rm x}$. Chemical potential $\mu$ is determined by
the ratio of the exciton number and the total spin: $N_{\rm x}(\mu)=\nu N_\phi/2\!-\!S$.
Calculating the quantity $\left.N_{\rm x}^{(0)}\!=\!N_{\rm x}\right|_{\mu=0}$, one obtains the
equilibrium number of excitons. We will find the rate at the final stage of the relaxation
process where $N_{\rm x}\!-\!N_{\rm x}^{(0)}\!\ll\! N_{\rm x}^{(0)}$. So, by employing the
quadratic approximation for the SE  kinetic energy, and changing in Eqs. (11)-(12) from
summations to integrations we obtain $R\!=\!\left(N_{\rm x}\!-\!N_{\rm
x}^{(0)}\right)/\tau_{{\rm srp}}$, where
$$
  1/\tau_{{\rm srp}}\!=\!\frac{(u^2\!+\!v^2)M_{\rm x}^3}{2\nu\pi\hbar}\left(\frac{\Delta\Lambda
  T}{l_B}\right)^2\left(e^{-\epsilon_{\rm Z}/T}\!\!-\!e^{-2\epsilon_{\rm Z}/T}\right)F_{\rm SRP}(\Lambda^2M_{\rm
  x}T/l_B^2,\,\epsilon_{\rm Z}/T)\,. \eqno (13)
$$
Here $F_{\rm SRP}(\alpha,\,\beta)$ is a dimensionless function
arising as a result of integrations over $q_1$ and $q_2$ and
averaging over angles $\theta_1\!=\!{\bf q}_1\land{\bf q}'$ and
$\theta_2\!=\!{\bf q}_2\land{\bf q}'$:
$$
  \begin{array}{r}
  \displaystyle{F_{\rm SRP}(\alpha,\,\beta)=\int_0^\infty\!\!{}\!{}\!\!\int_0^\infty \!\!\frac{e^{-x-y}\;dxdy}{(1\!-\!e^{-x\!-\!\beta})
  (1\!-\!e^{-y\!-\!\beta})(1\!-\!e^{-x\!-\!y\!-\!2\beta})}{}\qquad{}\qquad{}\qquad{}}\vspace{3mm}\\
  \displaystyle{\times\int_{-\pi}^\pi \!\!d\theta_1\!\int_{-\pi}^\pi
  \!\!d\theta_2\,r(x,y,\theta_1,\theta_2)\exp{[-\alpha\,
  r(x,y,\theta_1,\theta_2)]}}\,,
  \end{array}
$$
where $
r(x,y,\theta_1,\theta_2)\!=\!x\!+\!y\!+\!\beta/2\!-\!\sqrt{x\!+\!y\!+\!\beta}\,(\!\sqrt{x}\cos{\theta_1}\!+\!
\sqrt{y}\cos{\theta_2})\!+\!\sqrt{xy}\cos{(\theta_1\!-\!\theta_2)}$.

\vspace{3mm}

{\bf 4.} Now we calculate the numerical value of $1/\tau_{{\rm srp}}$ at typical SRP parameters
and compare it with inverse relaxation times $1/\tau_{e\!-e}$ and $1/\tau_{ph}$ governed by the
inter-SEs' interaction mechanism$\,$\cite{di99} and the SE-acoustic-phonon coupling.\cite{di96}
We carry out this analysis for the $\nu\!=\!1$ QHF  assuming that $\Delta\!=\!10\,$K and
$\Lambda\!=\!40\,$nm. The Zeeman splitting at $g\!=\!-0.44$ is $\epsilon_{\rm Z}\!=\!0.295B\,$K
($B$ is everywhere in Teslas), and the combination of SO parameters is estimated as
$u^2\!+\!v^2\!=\!10^{-3}/B$. The SE mass $M_{\rm x}$ might be calculated theoretically by using
general expressions for ${\cal E}_q$.\cite{by81,lo93} Yet, the result depends on specific
formfactor inherent in a given heterostructure due to finite thickness and it is therefore more
convenient to extract $M_{\rm x}$ immediately from experiments. According to recent data
available for currently used wide quantum wells,\cite{ga08,kuk06} we estimate that $1/M_{\rm
x}\!=\!9.24\sqrt{B}\,\;$K. Using Eq. (13), we thus calculate $1/\tau_{{\rm srp}}$ as a function
of temperature $T$ at given field $B$. The results are presented in Fig. 1 by dash curves. The
dot and dash-dot curves correspond to the $1/\tau_{e-e}$ and $1/\tau_{ph}$ values given by
formulas$\,$\cite{foot}
$$
1/\tau_{e\!-e}=\frac{2}{\hbar}(u^2+v^2)T\left(e^{-\epsilon_{\rm Z}/T}\!-\!e^{-2\epsilon_{\rm
Z}/T}\right)
  F_{e-e}(\epsilon_{\rm Z}/T) \,,  \eqno     (14)
$$
where
$$
  F_{e\!-e}(\beta)=
  \lefteqn{\int}\!\!\!\int\limits_{xy>\beta^2/4}^{}\!\!{}\frac{dx dy
  \quad (x+y+\beta)e^{-x\!-\!y}}
  {\left(xy\!-\!\beta^2/4\right)^{1/2}
  \left(1-e^{-\beta\!-\!x}\right)\left(1-e^{-\beta\!-\!y}\right)
  \left(1-e^{-2\beta\!-\!x\!-\!\beta}\right)}\,;
$$
and
$$
\tau^{-1}_{ph}=\frac{MT\epsilon_{\rm Z}(u^2+v^2)}{\hbar c_s p_0^3\l_B^2}
\left[\frac{\gamma_1(\epsilon_{\rm Z}/T)}{\tau_D}+10\frac{MT}{\tau_P} {}\!\left( \frac{\hbar
c_s}{\epsilon_{\rm Z}}\right)^4 \!\left(\frac{p_0}{\l_B}\right)^2 \!\!\gamma_2(\epsilon_{\rm
Z}/T)\right],\eqno (15)
$$
where
$$
\gamma_k(\beta) = (e^{2\beta}\!-\!e^{\beta})\!\!\int_0^{\infty}
      \!\!\!\frac{e^x x^k dx}{(e^{\beta + x} - 1)^2}\,,\quad k\!=\!1,\,2\,.
$$
(See Ref. \onlinecite{di96}; the used material parameters
characterizing the electron-phonon coupling are
$c_s\!=\!5.14\!\cdot\!10^{5}\,$cm/s,
$\tau_D\!=\!0.8\!\cdot\!10^{-12}\,$s${}^{-1}$,
$\tau_P\!=\!35\!\cdot\!10^{-12}\,$s${}^{-1}$, and $p_0 =
2.52\!\cdot\!10^{6}\,$cm${}^{-1}$; both kinds of $e$-$ph$
interaction, deformation and polarization ones, are taken into
account.)

It is seen from Fig. 1 that the SRP relaxation channel actually competes with other mechanisms
in the experimentally relevant range of parameters: namely, at fields $B\,\leq\, 5\,$ and
temperatures $T\sim 0.3-0.5\,$K. We have indicated above that the basic advantage of the SRP
channel, as compared to the $e$-$e$ one, consist in the absence of momentum conservation in the
coalescence process. On the other hand, the SRP mechanisms is also determined by effective SE-SE
collisions. Therefore the inverse relaxation time is proportional to the SE concentration and
drops exponentially as $\sim\!\exp{(-\epsilon_{\rm Z}/T)}$ with vanishing $T$ [rather than as
$\sim\!\exp{(-2\epsilon_{\rm Z}/T)}$ which occurs for the $e$-$e$ mechanism due to the SEs
momentum conservation!]. The phonon mechanism of SWR dominates at low temperatures due to its
weak temperature dependence ($\sim\!T$), in spite of small value of the electron-phonon coupling
constant in GaAs. The dependence on the filling factor in the case of integer QHF is only
determined by the SE mass $M_{\rm x}$ because $\nu$ in Eq. (13) is formally set equal to unit.
For fractional QHF there are both direct and indirect (through the mass $M_{\rm x}$) dependences
on $\nu$.

Finally we calculate the combined inverse relaxation time determined by the SO interaction:
$$
   1/\tau_{\rm tot}=1/\tau_{{\rm srp}}\!+\!1/\tau_{e\!-e}\!+\!1/\tau_{ph}\, \eqno (16)
$$
The result is presented by solid curves in Fig. 1. It is worth
mentioning that it demonstrates a good agreement with the measured
value $\tau_{\rm tot}\!\simeq 10\,$ns of Ref. \onlinecite{zh93}
when calculated for parameters $B$ and $T$ corresponding to the
experiment.

The authors acknowledge support of the RFBR and hospitality of the
Max Planck Institute for Physics of Complex Systems (Dresden)
where this work was partly carried out. The authors also thank
S.V. Iordanskii and L.V. Kulik for discussion.

\begin{figure}[h]
\begin{center} \vspace{-4.mm}
\includegraphics*[width=0.9\textwidth]{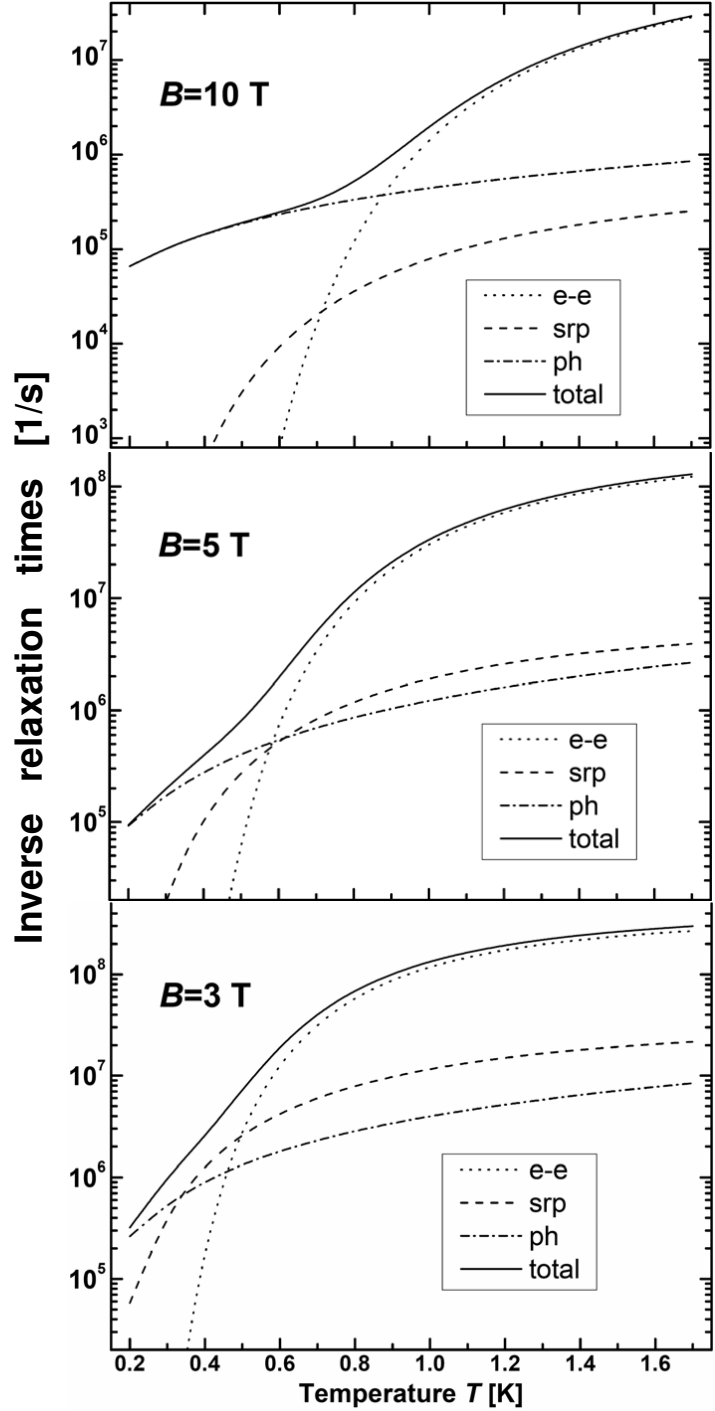}
\end{center}
\vspace{-10.mm}
 \caption{Inverse SWR times against $T$ calculated by using formulas (13)-(15) at
 $B\!=\!3,\,5,\,10\,$T. Specific material parameters are given in the text. Dash, dot, and dash-dot lines
 are for $1/\tau_{\rm srp}$, $1/\tau_{e\!-e}$ and $1/\tau_{ph}$, respectively. Solid lines present the result of
 calculation of the combined inverse time (16).}
\end{figure}

\begin{thebibliography}{99}
\bibitem{foot}
{The corresponding formula for inverse relaxation time in Ref.
\onlinecite{di99} contains a misprint. Now we present the
corrected result in Eq. (14).}
\end{thebibliography}
\end{document}